\documentclass[graybox]{svmult}

\usepackage{mathptmx}       % selects Times Roman as basic font
\usepackage{helvet}         % selects Helvetica as sans-serif font
\usepackage{courier}        % selects Courier as typewriter font
\usepackage{type1cm}        % activate if the above 3 fonts are
\usepackage{makeidx}         % allows index generation
\usepackage{graphicx}        % standard LaTeX graphics tool
\usepackage{multicol}        % used for the two-column index
\usepackage[bottom]{footmisc}% places footnotes at page bottom

\usepackage{bm}
\usepackage{latexsym}
\usepackage{amssymb}
\usepackage{dsfont}
\usepackage{amsmath}

\newcommand{\ud}{\mathrm{d}}

\newcommand{\uvec}[1]{\vec{#1}}
\newcommand{\pure}{\text{pure}}
\newcommand{\phys}{\text{phys}}

\makeindex             % used for the subject index
\begin{document}

\title*{Exploring the proton spin structure
}
% Use \titlerunning{Short Title} for an abbreviated version of
% your contribution title if the original one is too long
\author{C\'edric Lorc\'e}
% Use \authorrunning{Short Title} for an abbreviated version of
% your contribution title if the original one is too long
\institute{C\'edric Lorc\'e \at SLAC National Accelerator Laboratory, Stanford University, Menlo Park, CA 94025, USA, \email{clorce@slac.stanford.edu}\and
 IFPA,  AGO Department, Universit\'e de Li\` ege, Sart-Tilman, 4000 Li\`ege, Belgium, \email{C.Lorce@ulg.ac.be}}
%
% Use the package "url.sty" to avoid
% problems with special characters
% used in your e-mail or web address
%
\maketitle

%\abstract*{Each chapter should be preceded by an abstract (10--15 lines long) that summarizes the content. The abstract will appear \textit{online} at \url{www.SpringerLink.com} and be available with unrestricted access. This allows unregistered users to read the abstract as a teaser for the complete chapter. As a general rule the abstracts will not appear in the printed version of your book unless it is the style of your particular book or that of the series to which your book belongs.
%Please use the 'starred' version of the new Springer \texttt{abstract} command for typesetting the text of the online abstracts (cf. source file of this chapter template \texttt{abstract}) and include them with the source files of your manuscript. Use the plain \texttt{abstract} command if the abstract is also to appear in the printed version of the book.}

\abstract{Understanding the spin structure of the proton is one of the main challenges in hadronic physics. While the concepts of spin and orbital angular momentum are pretty clear in the context of non-relativistic quantum mechanics, the generalization of these concepts to quantum field theory encounters serious difficulties. It is however possible to define meaningful decompositions of the proton spin that are (in principle) measurable. We propose a summary of the present situation including recent developments and prospects of future developments.
}

\section{Introduction}	

Understanding how the proton spin arises from the spin and orbital motion of its constituents is one the most challenging key questions in hadronic physics. Hadrons are very peculiar physical systems as their constituents are highly relativistic and confined. One has therefore to use cunning in order to unravel their internal structure.

While it is clear that a proton at rest has total angular momentum $J=1/2$, the decomposition of the latter in terms of spin and orbital contributions associated with quarks and gluons is not unique, creating some confusion and raising serious controversies among physicists. Most of the discussions focused on determining which one of the proposed decompositions has to be considered as the ``physical'' or fundamental one. Now that the dust has settled, it turns out that the angular momentum decomposition is intrinsically ambiguous because of Lorentz and gauge symmetry. However, this does not imply that the question of the angular momentum decomposition does not make sense at all, but rather emphasizes the fact that the description of a physical phenomenon does not need to be unique. What is considered as the ``physical'' or fundamental description usually turns out to be the simplest or most convenient description at hand. 

While measurable quantities are necessarily gauge invariant, it has recently been recognized that they need not be local or \emph{manifestly} Lorentz covariant. Departing from locality or manifest Lorentz covariance leads to ambiguities as there exists in principle infinitely many ways to do so. What saves the day is that it is the way the physical system is probed, \emph{i.e.} the experimental configuration, which determines the natural or sensible departure from locality or manifest Lorentz covariance. For example, the internal structure of the proton is essentially probed in high-energy experiments which provide us with a natural preferred direction breaking manifest Lorentz covariance. This preferred direction can then be used to define the natural angular momentum decomposition.

One of the crucial questions now is to identify the experimental observables from which the orbital angular momentum (OAM) can be extracted. Many different relations and sum rules have been proposed in the last two decades, creating some sort of confusion. One of the remaining tasks consists in clarifying the validity and scope of these relations and sum rules.

In this contribution, we summarize the present situation and mention some recent developments. In section \ref{sec2}, we briefly discuss the two families of proton spin decompositions. In section \ref{sec3}, we collect various spin sum rules and relations. In section \ref{sec4}, we introduce the notion of quark spin-orbit correlation and show how it is related to measurable parton distributions. Finally, we collect our conclusions in section \ref{sec5}. For the interested reader, more detailed discussions can be found in the recent reviews~\cite{Leader:2013jra,Wakamatsu:2014zza}.

\section{Kinetic and canonical spin decompositions}\label{sec2}

\begin{figure}
\centering
  \includegraphics[width=8cm]{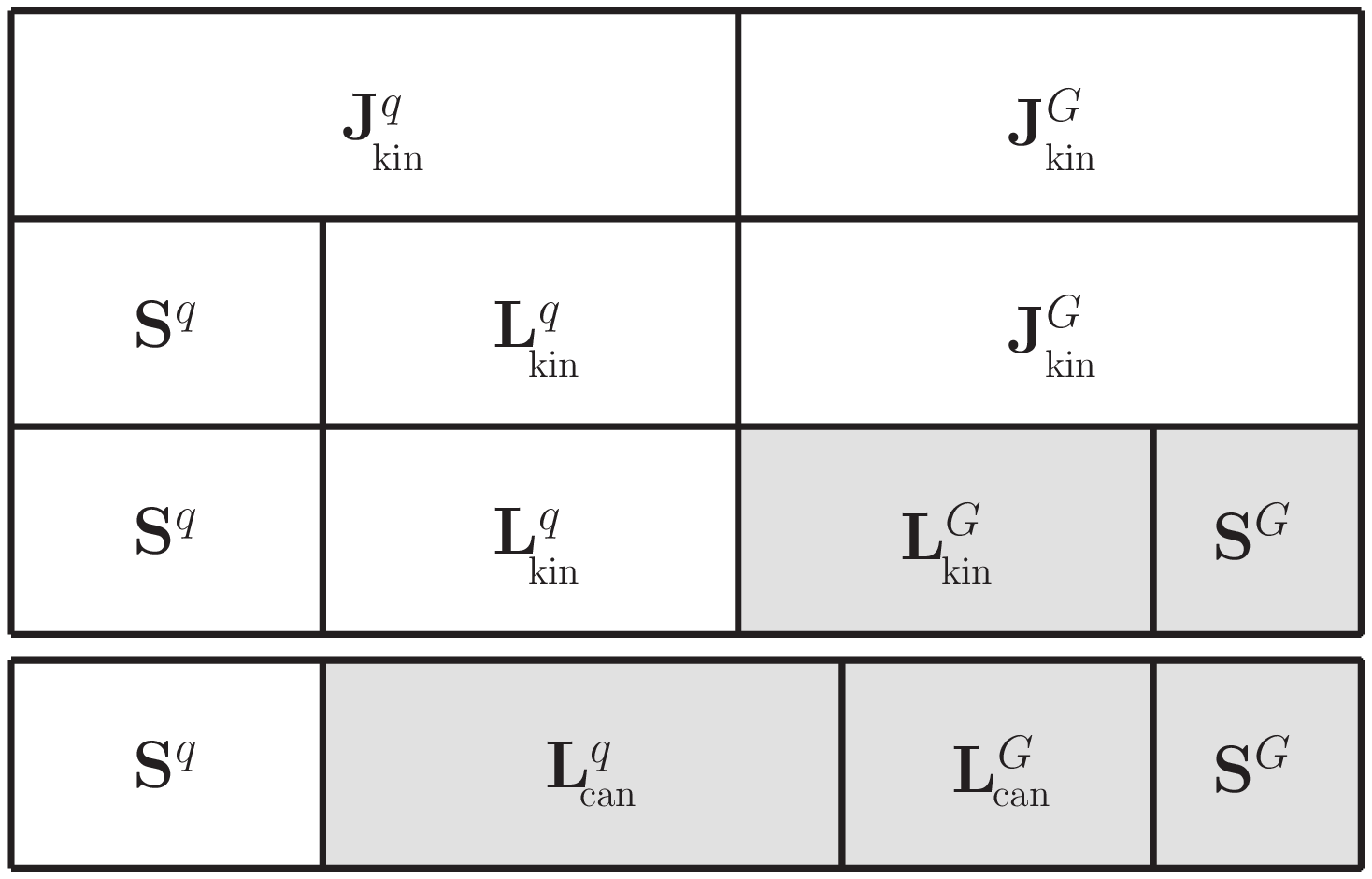}
\caption{The proton spin decompositions. The first three are the Belinfante, Ji and Wakamatsu versions of the kinetic decomposition. The last one is the Chen \emph{et al.} canonical decomposition which is the manifestly gauge-invariant version of the Jaffe-Manohar decomposition. The contributions in gray are nonlocal or not manifestly Lorentz covariant.}
\label{fig:1}       
\end{figure}

There are essentially two types of decompositions of the proton spin operator: kinetic (also known as mechanical) and canonical. These two types differ by how the OAM operator is split into the quark ($q$) and gluon ($G$) contributions
\begin{equation}\label{AMdec}
\begin{aligned}
\uvec J&=\uvec S^q+\uvec L^q_\text{kin}+\uvec L^G_\text{kin}+\uvec S^G,\\
&=\uvec S^q+\uvec L^q_\text{can}+\uvec L^G_\text{can}+\uvec S^G,
\end{aligned}
\end{equation}
where
\begin{equation}
\begin{aligned}
\uvec S^q&=\int\ud^3r\,\psi^\dag\tfrac{1}{2}\bm\Sigma\psi, \qquad&\uvec S^G&=\int\ud^3r\,\uvec E^a\times\uvec A^a_\phys,\\
\uvec L^q_\text{kin}&=\int\ud^3r\,\psi^\dag(\uvec r\times i\uvec D)\psi,\qquad&\uvec L^G_\text{kin}&=\uvec L^G_\text{can}-\int\ud^3r\,(\uvec D^{ab}\cdot\uvec E^b)\,\uvec x\times\uvec A^a_\phys,\\
\uvec L^q_\text{can}&=\int\ud^3r\,\psi^\dag(\uvec r\times i\uvec D_\pure)\psi,&\uvec L^G_\text{can}&=-\int\ud^3r\,E^{aj}(\uvec r\times\uvec D^{ab}_\pure) A^{bj}_\phys.
\end{aligned}
\end{equation}
The gauge field has been decomposed into two parts $\uvec A=\uvec A_\pure+\uvec A_\phys$ where $\uvec A_\pure$ is a pure-gauge potential. The pure-gauge covariant derivatives are then given by $\uvec D_\pure=-\bm\nabla-ig\uvec A_\pure$ and $\uvec D^{ab}_\pure=-\bm\nabla\delta^{ab}-gf^{abc}\uvec A^c_\pure$. A nice physical interpretation of the difference $\uvec L_\text{pot}=\uvec L^q_\text{kin}-\uvec L^q_\text{can}=\uvec L^G_\text{can}-\uvec L^G_\text{kin}$, known as the potential OAM, has been proposed in Ref.~\cite{Burkardt:2012sd}.

The complete gauge-invariant kinetic and canonical decompositions~\eqref{AMdec} are known in the literature as the Wakamatsu~\cite{Wakamatsu:2010qj,Wakamatsu:2010cb} and Chen \emph{et al.}~\cite{Chen:2008ag,Chen:2009mr} decompositions, respectively. The Chen \emph{et al.} decompositon can be seen as a gauge-invariant version (or extension) of the Jaffe-Manohar decomposition~\cite{Jaffe:1989jz}. These complete gauge-invariant decompositions seem to contradict textbook claims about the impossibility of separating in a gauge-invariant way the gluon angular momentum into spin and OAM contributions. This impossibility is circumvented by introducing the non-local fields $\uvec A_\pure$ and $\uvec A_\phys$~\cite{Lorce:2012rr,Lorce:2012ce}, where the pure-gauge field $\uvec A_\pure$ plays the role of a background field~\cite{Lorce:2013gxa,Lorce:2013bja}. Background dependence then implies that the decomposition $\uvec A=\uvec A_\pure+\uvec A_\phys$ comes with a new freedom
\begin{equation}\label{Stueckelberg}
\uvec A_\pure\mapsto\uvec A_\pure+\uvec B,\qquad\uvec A_\phys\mapsto\uvec A_\phys-\uvec B,
\end{equation}
referred to as the Stueckelberg symmetry~\cite{Lorce:2012rr,Stoilov:2010pv}, making the decompositions ambiguous as \emph{a priori} any pure-gauge field $\uvec A_\pure$ can be used. This issue is however solved by noting that the actual experimental conditions determine the form of the background field to be used~\cite{Lorce:2012rr,Wakamatsu:2014toa}.

Incomplete kinetic decompositions avoid the uniqueness issue from the beginning. In the Ji decomposition~\cite{Ji:1996ek}, the gluon spin and OAM contributions are combined to form the gluon total angular momentum 
\begin{equation}
\uvec J^G_\text{kin}=\uvec S^G+\uvec L^G_\text{kin}=\int\ud^3r\,\uvec r\times(\uvec E^a\times\uvec B^a)
\end{equation}
which is local and therefore free from the Stueckelberg ambiguity. In the Belinfante decomposition, one further combines the quark spin and OAM contributions into the quark total angular momentum
\begin{equation}
\uvec J^q_\text{kin}=\uvec S^q+\uvec L^q_\text{kin}=\int\ud^3r\,\overline\psi\,\uvec r\times(\gamma^0\,\tfrac{i}{2}\uvec D+\bm\gamma\,\tfrac{i}{2}D^0)\psi
\end{equation}
so that one can write $\uvec J^{q,G}_\text{kin}=\uvec r\times\uvec P^{q,G}_\text{kin}$ with $P^j_\text{kin}=T^{0j}_\text{kin}=T^{j0}_\text{kin}$ where $T^{\mu\nu}_\text{kin}$ is the symmetric kinetic (or Belinfante-Rosenfeld) energy-momentum tensor. See Fig.~\ref{fig:1} for a summary of the decompositions.

\section{Spin sum rules and relations}\label{sec3}

Using the Belinfante-Rosenfeld version of the energy-momentum tensor, Ji obtained the remarkable result that the quark/gluon total kinetic angular momentum can be expressed in terms of twist-2 generalized parton distributions (GPDs)~\cite{Ji:1996ek}
\begin{equation}\label{Jirel}
\langle J^{q,G}_\text{kin}\rangle=\tfrac{1}{2}\int\ud x\,x[H^{q,G}(x,0,0)+E^{q,G}(x,0,0)].
\end{equation}
This relation holds for the longitudinal component $J_L=\uvec J\cdot\uvec P/|\uvec P|$ and does not depend on the magnitude of the proton momentum $|\uvec P|$~\cite{Ji:1997pf,Leader:2011cr,Leader:2012md}. By rotational symmetry, it is also valid for the transverse component, but only in the proton rest frame. Considering the transverse component of the Pauli-Lubanski vector does not prevent frame dependence of the separate quark and gluon contributions~\cite{Leader:2012ar,Leader:2013jra,Hatta:2012jm,Harindranath:2013goa}. 

Subtracting from Eq.~\eqref{Jirel} the longitudinal quark spin contribution, which is given by the isoscalar axial-vector form factor (FF) in the $\overline{MS}$ scheme
\begin{equation}
\langle S^q\rangle=\tfrac{1}{2}\,G^q_A(0),
\end{equation}
one obtains the following expression for the longitudinal quark kinetic OAM 
\begin{equation}\label{OAMeq}
\langle L^q_\text{kin}\rangle=\tfrac{1}{2}\int\ud x\,x[H^q(x,0,0)+E^q(x,0,0)]-\tfrac{1}{2}\,G^q_A(0).
\end{equation}
The same quantitie can also be expressed in terms of a twist-3 GPD~\cite{Penttinen:2000dg,Kiptily:2002nx,Hatta:2012cs,Lorce:2015lna}
\begin{equation}
\langle L^q_\text{kin}\rangle=-\int\ud x\,xG^q_2(x,0,0)
\end{equation}
which appears in the longitudinal target spin asymmetry of deeply virtual Compton scattering~\cite{Courtoy:2013oaa}.

The most intuitive expression for OAM is as a phase-space integral~\cite{Lorce:2011kd,Lorce:2011ni}
\begin{equation}\label{OAMWigner}
\langle L^{q,G}(\mathcal W)\rangle=\int\ud x\,\ud^2k_\perp\,\ud^2b_\perp\,(\vec b_\perp\times\vec k_\perp)_z\,\rho^{q,G}_{++}(x,\vec k_\perp,\vec b_\perp;\mathcal W),
\end{equation}
where the relativistic phase-space or Wigner distribution $\rho^{q,G}_{++}(x,\vec k_\perp,\vec b_\perp;\mathcal W)$ can be interpreted as giving the (quasi-)probability for finding an unpolarized quark/gluon with momentum $(xP^+,\uvec k_\perp)$ at the transverse position $\uvec b_\perp$ inside a longitudinally polarized proton. In this semi-classical interpretation, the Euclidean subgroup of the light-front formalism plays a crucial role in providing a well-defined transverse center of the proton~\cite{Soper:1976jc,Burkardt:2000za,Burkardt:2005hp}. These phase-space distributions are related by Fourier transform to the so-called generalized transverse-momentum dependent distributions (GTMDs)~\cite{Meissner:2009ww,Lorce:2011dv,Lorce:2013pza}, leading to the simple relation~\cite{Lorce:2011kd,Hatta:2011ku,Kanazawa:2014nha}
\begin{equation}\label{LzGTMD}
\langle L^{q,G}(\mathcal W)\rangle=-\int\ud x\,\ud^2k_\perp\,\tfrac{\uvec k^2_\perp}{M^2}\,F^{q,G}_{14}(x,0,\uvec k_\perp,\uvec 0_\perp;\mathcal W).
\end{equation}
The type of OAM is determined by the shape of the Wilson line $\mathcal W$, namely $\langle L^{q,G}_\text{kin}\rangle=\langle L^{q,G}(\mathcal W_\text{straight})\rangle$ and $\langle L^{q,G}_\text{can}\rangle=\langle L^{q,G}(\mathcal W_\text{staple})\rangle$~\cite{Burkardt:2012sd,Lorce:2012ce,Ji:2012sj}. Unfortunately, it is not known so far how to extract GTMDs from actual experiments, except perhaps at small $x$~\cite{Meissner:2009ww}. Interestingly, they are however in principle calculable on the lattice~\cite{Ji:2013dva}.

In the context of quark models, it has also been suggested that the quark canonical OAM could be related to a transverse-momentum dependent distribution (TMD) 
\begin{equation}\label{pretzelosity}
\langle L_\text{can}^q\rangle=-\int\ud x\,\ud^2k_\perp\,\tfrac{\uvec k_\perp^2}{2M^2}\,h_{1T}^{\perp q}(x,\uvec k^2_\perp),
\end{equation}
but this relation is not valid in general in QCD~\cite{Lorce:2011kn} just like other relations among the TMDs~\cite{Lorce:2011zta}. In Table~\ref{OAMtable}, the various expressions~\eqref{OAMeq},~\eqref{LzGTMD} and~\eqref{pretzelosity} for the quark OAM are compared in two light-front quark models: the light-front constituent quark model (LFCQM)~\cite{Boffi:2002yy,Boffi:2003yj,Pasquini:2005dk,Pasquini:2006iv,Pasquini:2008ax} and the light-front chiral quark-soliton model (LF$\chi$QSM)~\cite{Lorce:2006nq,Lorce:2007as,Lorce:2007fa}. While all the expressions agree for the total OAM, as they should, they differ in the flavor decomposition.

\begin{table}[t]
\begin{center}
\caption{\footnotesize{Results for the quark OAM from two light-front quark models for $u$, $d$ and total ($u+d$) quark contributions.}}\label{OAMtable}
\begin{tabular}{@{\quad}c@{\quad}c@{\quad}|@{\quad}c@{\quad}c@{\quad}c@{\quad}|@{\quad}c@{\quad}c@{\quad}c@{\quad}}\hline\noalign{\smallskip}
\multicolumn{2}{@{\quad}c@{\quad}|@{\quad}}{Model}&\multicolumn{3}{c@{\quad}|@{\quad}}{LFCQM}&\multicolumn{3}{c@{\quad}}{LF$\chi$QSM}\\
\multicolumn{2}{@{\quad}c@{\quad}|@{\quad}}{$q$}&$u$&$d$&Total&$u$&$d$&Total\\
\noalign{\smallskip}\svhline\noalign{\smallskip}
$\langle L^q_\text{kin}\rangle$&Eq.~\eqref{OAMeq}&$0.071$&$0.055$&$0.126$&$-0.008$&$0.077$&$0.069$\\
$\langle L^q_\text{can}\rangle$&Eq.~\eqref{LzGTMD}&$0.131$&$-0.005$&$0.126$&$0.073$&$-0.004$&$0.069$\\

$\langle L^q_\text{can}\rangle$&Eq.~\eqref{pretzelosity}&$0.169$&$-0.042$&$0.126$&$0.093$&$-0.023$&$0.069$\\
\noalign{\smallskip}\hline\noalign{\smallskip}
\end{tabular}
\end{center}
\end{table}

\section{Spin-orbit correlation}\label{sec4}

What is referred to as the quark spin/OAM contribution to the proton spin corresponds more precisely to the \emph{correlation} between the quark spin/OAM and the proton spin. There exists another interesting independent correlation characterizing the proton spin structure, although it does not appear in the proton spin decomposition, namely the correlation between the quark spin and the quark OAM. Like the OAM operators, one can define a kinetic and a canonical version of this spin-orbit correlation~\cite{Lorce:2011kd,Lorce:2014mxa}
\begin{equation}
\begin{aligned}
\uvec C^q_\text{kin}&=\int\ud^3x\,\psi^\dag\gamma_5(\uvec x\times i\uvec D)\psi,\\
\uvec C^q_\text{can}&=\int\ud^3x\,\psi^\dag\gamma_5(\uvec x\times i\uvec D_\pure)\psi.
\end{aligned}
\end{equation}

Like the average kinetic OAM contribution to the proton spin, the average quark longitudinal spin-orbit correlation can be expressed in terms of twist-2 and twist-3 GPDs~\cite{Lorce:2014mxa}
\begin{equation}\label{SOtwist2}
\begin{aligned}
\langle C^q_\text{kin}\rangle&=\tfrac{1}{2}\int\ud x\,x\tilde H^q(x,0,0)-\tfrac{1}{2}\,[F^q_1(0)-\tfrac{m_q}{2M_N}\,H^q_1(0)],\\
&=-\int\ud x\,x[\tilde G^q_2(x,0,0)+2\tilde G^q_4(x,0,0)].
\end{aligned}
\end{equation}
Remarkably, this shows that not only the first moment but also the second moment of the quark helicity distribution has physical interest.

The quark spin-orbit correlation can naturally also be expressed as a phase-space integral~\cite{Lorce:2011kd,Lorce:2014mxa}
\begin{equation}\label{OAMWigner}
\langle C^q(\mathcal W)\rangle=\int\ud x\,\ud^2k_\perp\,\ud^2b_\perp\,(\vec b_\perp\times\vec k_\perp)_z\,\rho^{[\gamma^+\gamma_5]q}_{++}(x,\vec k_\perp,\vec b_\perp;\mathcal W),
\end{equation}
where the relativistic phase-space distribution $\rho^{[\gamma^+\gamma_5]q}_{++}(x,\vec k_\perp,\vec b_\perp;\mathcal W)$ can be interpreted as giving the difference between the (quasi-)probabilistic distributions of quarks with polarization parallel and antiparallel to the longitudinal direction. In terms of the GTMDs, this relation reads~\cite{Lorce:2011kd,Kanazawa:2014nha,Lorce:2014mxa}
\begin{equation}
\langle C^q(\mathcal W)\rangle=\int\ud x\,\ud^2k_\perp\,\tfrac{\uvec k^2_\perp}{M^2}\,G^q_{11}(x,0,\uvec k_\perp,\uvec 0_\perp;\mathcal W).
\end{equation}
Once again, the shape of the Wilson line $\mathcal W$ determines the type of spin-orbit correlation, namely $\langle C^q_\text{kin}\rangle=\langle C^q(\mathcal W_\text{straight})\rangle$ and $\langle C^q_\text{can}\rangle=\langle C^q(\mathcal W_\text{staple})\rangle$~\cite{Lorce:2014mxa}.

Because of the valence number constraints $F^u_1(0)=2$ and $F^d_1(0)=1$ and the small mass ratio $m_{u,d}/4M_N\sim 10^{-3}$, the essential non-perturbative input we need is the second moment of the quark helicity distribution
\begin{equation}
\int_{-1}^1\ud x\,x\tilde H^q(x,0,0)=\int_0^1\ud x\,x[\Delta q(x)-\Delta\overline q(x)].
\end{equation}
Contrary to the lowest moment $\int_{-1}^1\ud x\,\tilde H_q(x,0,0)=\int_0^1\ud x\,[\Delta q(x)+\Delta\overline q(x)]$, this second moment cannot simply be extracted from deep-inelastic scattering (DIS) polarized data. However, by combining inclusive and semi-inclusive DIS, separate quark and antiquark contributions can be extracted~\cite{Leader:2010rb}. They can also be computed on the lattice~\cite{Bratt:2010jn}. In Table~\ref{Modelresults}, the first two moments of the quark helicity distributions computed within the naive quark model (NQM), the LFCQM and the LF$\chi$QSM are compared with the values obtained from inclusive and semi-inclusive DIS data~\cite{Leader:2010rb} and from lattice calculations~\cite{Bratt:2010jn}.

\begin{table}[t]
\caption{Comparison between the lowest two axial moments for $u$ and $d$ quarks as predicted by various quark models, with the corresponding values obtained from the LSS fit to experimental data at $\mu^2=1$ GeV$^2$ and lattice calculations at $\mu^2=4$ GeV$^2$ and pion mass $m_\pi=293$ MeV.}
\centering\label{Modelresults}
\begin{tabular}{@{\quad}c@{\quad}|@{\quad}c@{\quad}c@{\quad}|@{\quad}c@{\quad}c@{\quad}}\hline\noalign{\smallskip}
Model &$\int\ud x\,\tilde H^u(x,0,0)$&$\int\ud x\,\tilde H^d(x,0,0)$&$\int\ud x\,x\tilde H^u(x,0,0)$&$\int\ud x\,x\tilde H^d(x,0,0)$\\
\noalign{\smallskip}\svhline\noalign{\smallskip}
NQM&$4/3$&$-1/3$&$4/9$&$-1/9$\\
LFCQM&$0.995$&$-0.249$&$0.345$&$-0.086$\\
LF$\chi$QSM&$1.148$&$-0.287$&$0.392$&$-0.098$\\
\noalign{\smallskip}\svhline\noalign{\smallskip}
LSS~\cite{Leader:2010rb}&$0.82$&$-0.45$&$\approx 0.19$&$\approx -0.06$\\
Lattice~\cite{Bratt:2010jn}&$0.82(7)$&$-0.41(7)$&$\approx 0.20$&$\approx -0.05$\\
\noalign{\smallskip}\hline\noalign{\smallskip}
\end{tabular}
\end{table}

From these estimates, one obtains a negative kinetic quark spin-orbit correlation for both quark flavors, $\langle C^u_\text{kin}\rangle\approx -0.8$ and $\langle C^d_\text{kin}\rangle\approx -0.55$, meaning that in average the quark spin and kinetic OAM are expected to be antiparallel. On the contrary, the canonical version of the quark spin-orbit correlation appears to be positive in the models~\cite{Lorce:2011kd}, showing the importance of the quark-gluon interaction.

\section{Conclusion}\label{sec5}

There are essentially two types of proton spin decompositions: the kinetic one and the canonical one. It has recently been recognized that both are interesting and in principle measurable. The crucial missing piece in the proton spin decomposition is the contribution coming from the quark and gluon orbital angular momentum. Several relations and sum rules have been proposed in the literature, but few proved to be of practical significance. The current most promising approaches are based on the extraction of generalized parton distributions at twist 2 and 3 from experiments, and the direct calculation of orbital angular momentum on the lattice.

Another important aspect of the proton spin structure is the spin-orbit correlation which escaped attention until recently because it does not contribute to the proton spin decomposition. Like the orbital angular momentum, there are two types of spin-orbit correlations, and both are in principle measurable. This piece of information is of crucial importance if one aims at obtaining a complete description of the proton spin structure.  

\begin{acknowledgement}
I benefited a lot from many discussions and collaborations with E. Leader, B. Pasquini and  M. Wakamatsu. This work was supported by the Belgian Fund F.R.S.-FNRS \emph{via} the contract of Charg\'e de Recherches.
\end{acknowledgement}


\begin{thebibliography}{99.}%
% and use \bibitem to create references.
%
% Use the following syntax and markup for your references if 
% the subject of your book is from the field 
% "Mathematics, Physics, Statistics, Computer Science"
%
% Contribution 

% Monograph
%\bibitem{phys-mono} H. Ibach, H. L\"uth, \textit{Solid-State Physics}, 2nd edn. (Springer, New York, 1996), pp. 45-56 
%
% Journal article
%\bibitem{phys-journal} S. Preuss, A. Demchuk Jr., M. Stuke, Appl. Phys. A \textbf{61}

\bibitem{Boffi:2002yy}
  S.~Boffi, B.~Pasquini, and M.~Traini,
  %``Linking generalized parton distributions to constituent quark models,''
  Nucl.\ Phys.\  B {\bf 649}, 243 (2003).
%  [arXiv:hep-ph/0207340].
  %%CITATION = NUPHA,B649,243;%%

\bibitem{Boffi:2003yj}
S.~Boffi, B.~Pasquini, and M.~Traini,
  %``Helicity-dependent generalized parton distributions in constituent  quark
  %models,''
  Nucl.\ Phys.\  B {\bf 680}, 147 (2004).
%  [arXiv:hep-ph/0311016].
  %%CITATION = NUPHA,B680,147;%%

\bibitem{Bratt:2010jn} 
  J.~D.~Bratt {\it et al.}  [LHPC Collaboration],
  %``Nucleon structure from mixed action calculations using 2+1 flavors of asqtad sea and domain wall valence fermions,''
  Phys.\ Rev.\ D {\bf 82}, 094502 (2010).
  %[arXiv:1001.3620 [hep-lat]].
  %%CITATION = ARXIV:1001.3620;%%
  %131 citations counted in INSPIRE as of 22 Sep 2014

\bibitem{Burkardt:2000za} 
  M.~Burkardt,
  %``Impact parameter dependent parton distributions and off forward parton distributions for zeta ---> 0,''
  Phys.\ Rev.\ D {\bf 62}, 071503 (2000)
  [Erratum-ibid.\ D {\bf 66}, 119903 (2002)].
 % [hep-ph/0005108].
  %%CITATION = HEP-PH/0005108;%%
  %419 citations counted in INSPIRE as of 22 Sep 2014

\bibitem{Burkardt:2005hp} 
  M.~Burkardt,
  %``Transverse deformation of parton distributions and transversity decomposition of angular momentum,''
  Phys.\ Rev.\ D {\bf 72}, 094020 (2005).
 % [hep-ph/0505189].
  %%CITATION = HEP-PH/0505189;%%
  %127 citations counted in INSPIRE as of 22 Sep 2014

\bibitem{Burkardt:2012sd} 
  M.~Burkardt,
  %``Parton Orbital Angular Momentum and Final State Interactions,''
  Phys.\ Rev.\ D {\bf 88}, no. 1, 014014 (2013).
  %[arXiv:1205.2916 [hep-ph]].
  %%CITATION = ARXIV:1205.2916;%%
  %25 citations counted in INSPIRE as of 17 Sep 2014

\bibitem{Chen:2008ag} 
  X.~-S.~Chen, X.~-F.~Lu, W.~-M.~Sun, F.~Wang and T.~Goldman,
  %``Spin and orbital angular momentum in gauge theories: Nucleon spin structure and multipole radiation revisited,''
  Phys.\ Rev.\ Lett.\  {\bf 100}, 232002 (2008).
  %[arXiv:0806.3166 [hep-ph]].
  %%CITATION = ARXIV:0806.3166;%%
  %70 citations counted in INSPIRE as of 09 Jan 2014

\bibitem{Chen:2009mr} 
  X.~-S.~Chen, W.~-M.~Sun, X.~-F.~Lu, F.~Wang and T.~Goldman,
  %``Do gluons carry half of the nucleon momentum?,''
  Phys.\ Rev.\ Lett.\  {\bf 103}, 062001 (2009).
  %[arXiv:0904.0321 [hep-ph]].
  %%CITATION = ARXIV:0904.0321;%%
  %53 citations counted in INSPIRE as of 09 Jan 2014

\bibitem{Courtoy:2013oaa} 
  A.~Courtoy, G.~R.~Goldstein, J.~O.~G.~Hernandez, S.~Liuti and A.~Rajan,
  %``On the Observability of the Quark Orbital Angular Momentum Distribution,''
  Phys.\ Lett.\ B {\bf 731}, 141 (2014).
  %[arXiv:1310.5157 [hep-ph]].
  %%CITATION = ARXIV:1310.5157;%%
  %9 citations counted in INSPIRE as of 15 Apr 2015

\bibitem{Harindranath:2013goa} 
  A.~Harindranath, R.~Kundu and A.~Mukherjee,
  %``On transverse spin sum rules,''
  Phys.\ Lett.\ B {\bf 728}, 63 (2014).
  %[arXiv:1308.1519 [hep-ph]].
  %%CITATION = ARXIV:1308.1519;%%
  %2 citations counted in INSPIRE as of 17 Sep 2014

\bibitem{Hatta:2011ku} 
  Y.~Hatta,
  %``Notes on the orbital angular momentum of quarks in the nucleon,''
  Phys.\ Lett.\ B {\bf 708}, 186 (2012).
  %[arXiv:1111.3547 [hep-ph]].
  %%CITATION = ARXIV:1111.3547;%%
  %32 citations counted in INSPIRE as of 09 Jan 2014

\bibitem{Hatta:2012cs} 
  Y.~Hatta and S.~Yoshida,
  %``Twist analysis of the nucleon spin in QCD,''
  JHEP {\bf 1210}, 080 (2012).
  %[arXiv:1207.5332 [hep-ph]].
  %%CITATION = ARXIV:1207.5332;%%
  %15 citations counted in INSPIRE as of 22 Sep 2014

\bibitem{Hatta:2012jm} 
  Y.~Hatta, K.~Tanaka and S.~Yoshida,
  %``Twist-three relations of gluonic correlators for the transversely polarized nucleon,''
  JHEP {\bf 1302}, 003 (2013).
  %[arXiv:1211.2918 [hep-ph]].
  %%CITATION = ARXIV:1211.2918;%%
  %9 citations counted in INSPIRE as of 17 Sep 2014

\bibitem{Jaffe:1989jz} 
  R.~L.~Jaffe and A.~Manohar,
  %``The G(1) Problem: Fact and Fantasy on the Spin of the Proton,''
  Nucl.\ Phys.\ B {\bf 337}, 509 (1990).
  %%CITATION = NUPHA,B337,509;%%
  %547 citations counted in INSPIRE as of 09 Jan 2014

\bibitem{Ji:1996ek} 
  X.~-D.~Ji,
  %``Gauge-Invariant Decomposition of Nucleon Spin,''
  Phys.\ Rev.\ Lett.\  {\bf 78}, 610 (1997).
  %[hep-ph/9603249].
  %%CITATION = HEP-PH/9603249;%%
  %1133 citations counted in INSPIRE as of 09 Jan 2014

\bibitem{Ji:1997pf} 
  X.~D.~Ji,
  %``Lorentz symmetry and the internal structure of the nucleon,''
  Phys.\ Rev.\ D {\bf 58}, 056003 (1998).
  %[hep-ph/9710290].
  %%CITATION = HEP-PH/9710290;%%
  %38 citations counted in INSPIRE as of 15 Apr 2015

\bibitem{Ji:2013dva} 
  X.~Ji,
  %``Parton Physics on a Euclidean Lattice,''
  Phys.\ Rev.\ Lett.\  {\bf 110}, 262002 (2013).
  %[arXiv:1305.1539 [hep-ph]].
  %%CITATION = ARXIV:1305.1539;%%
  %27 citations counted in INSPIRE as of 17 Sep 2014

\bibitem{Ji:2012sj} 
  X.~Ji, X.~Xiong and F.~Yuan,
  %``Proton Spin Structure from Measurable Parton Distributions,''
  Phys.\ Rev.\ Lett.\  {\bf 109}, 152005 (2012).
 % [arXiv:1202.2843 [hep-ph]].
  %%CITATION = ARXIV:1202.2843;%%
  %46 citations counted in INSPIRE as of 26 Jun 2014

\bibitem{Kanazawa:2014nha} 
  K.~Kanazawa, C.~Lorc\'e, A.~Metz, B.~Pasquini and M.~Schlegel,
  %``Twist-2 Generalized TMDs and the Spin/Orbital Structure of the Nucleon,''
  Phys.\ Rev.\ D {\bf 90}, 014028 (2014).
  %[arXiv:1403.5226 [hep-ph]].
  %%CITATION = ARXIV:1403.5226;%%
  %4 citations counted in INSPIRE as of 17 Sep 2014

\bibitem{Kiptily:2002nx} 
  D.~V.~Kiptily and M.~V.~Polyakov,
  %``Genuine twist three contributions to the generalized parton distributions from instantons,''
  Eur.\ Phys.\ J.\ C {\bf 37}, 105 (2004).
  %[hep-ph/0212372].
  %%CITATION = HEP-PH/0212372;%%
  %20 citations counted in INSPIRE as of 22 Sep 2014

\bibitem{Leader:2011cr} 
  E.~Leader,
  %``New relation between transverse angular momentum and generalized parton distributions,''
  Phys.\ Rev.\ D {\bf 85}, 051501 (2012).
  %[arXiv:1109.1230 [hep-ph]].
  %%CITATION = ARXIV:1109.1230;%%
  %14 citations counted in INSPIRE as of 15 Apr 2015


\bibitem{Leader:2012ar} 
  E.~Leader,
  %``A critical assessment of the angular momentum sum rules,''
  Phys.\ Lett.\ B {\bf 720}, 120 (2013).
  %[arXiv:1211.3957 [hep-ph]].
  %%CITATION = ARXIV:1211.3957;%%
  %8 citations counted in INSPIRE as of 17 Sep 2014

\bibitem{Leader:2012md} 
  E.~Leader and C.~Lorc\'e,
  %``Comment on 'Proton Spin Structure from Measurable Parton Distributions' by Ji, Xiong and Yuan (PRL109, 152005 (2012)),''
  Phys.\ Rev.\ Lett.\  {\bf 111}, 039101 (2013).
  %[arXiv:1211.4731 [hep-ph]].
  %%CITATION = ARXIV:1211.4731;%%
  %9 citations counted in INSPIRE as of 17 Sep 2014

\bibitem{Leader:2013jra} 
  E.~Leader and C.~Lorc\'e,
  %``The angular momentum controversy: What's it all about and does it matter?,''
  Phys.\ Rept.\  {\bf 541}, 163 (2014).
  %[arXiv:1309.4235 [hep-ph]].
  %%CITATION = ARXIV:1309.4235;%%
  %27 citations counted in INSPIRE as of 17 Sep 2014

\bibitem{Leader:2010rb} 
  E.~Leader, A.~V.~Sidorov and D.~B.~Stamenov,
  %``Determination of Polarized PDFs from a QCD Analysis of Inclusive and Semi-inclusive Deep Inelastic Scattering Data,''
  Phys.\ Rev.\ D {\bf 82}, 114018 (2010).
  %[arXiv:1010.0574 [hep-ph]].
  %%CITATION = ARXIV:1010.0574;%%
  %74 citations counted in INSPIRE as of 22 Sep 2014

\bibitem{Lorce:2006nq}
  C.~Lorc\'e,
  %``Improvement of the Theta+ width estimation method on the light cone,''
  Phys.\ Rev.\  D {\bf 74}, 054019 (2006).
  %[arXiv:hep-ph/0603231].
  %%CITATION = PHRVA,D74,054019;%%

\bibitem{Lorce:2007as}
  C.~Lorc\'e,
  %``Baryon vector and axial content up to the 7Q component,''
  Phys.\ Rev.\  D {\bf 78}, 034001 (2008).
  %[arXiv:0708.3139 [hep-ph]].
  %%CITATION = PHRVA,D78,034001;%%
  
\bibitem{Lorce:2007fa}
  C.~Lorc\'e,
  %``Tensor charges of light baryons in the Infinite Momentum Frame,''
  Phys.\ Rev.\  D {\bf 79}, 074027 (2009). 
  %[arXiv:0708.4168 [hep-ph]].
  %%CITATION = PHRVA,D79,074027;%%

\bibitem{Lorce:2012rr} 
  C.~Lorc\'e,
  %``Geometrical approach to the proton spin decomposition,''
  Phys.\ Rev.\ D {\bf 87}, 034031 (2013).
  %[arXiv:1205.6483 [hep-ph]].
  %%CITATION = ARXIV:1205.6483;%%
  %25 citations counted in INSPIRE as of 09 Jan 2014

\bibitem{Lorce:2012ce} 
  C.~Lorc\'e,
  %``Wilson lines and orbital angular momentum,''
  Phys.\ Lett.\ B {\bf 719}, 185 (2013).
  %[arXiv:1210.2581 [hep-ph]].
  %%CITATION = ARXIV:1210.2581;%%
  %16 citations counted in INSPIRE as of 09 Jan 2014

\bibitem{Lorce:2013gxa} 
  C.~Lorc\'e,
  %``Gauge-covariant canonical formalism revisited with application to the proton spin decomposition,''
  Phys.\ Rev.\ D {\bf 88}, 044037 (2013).
  %[arXiv:1302.5515 [hep-ph]].
  %%CITATION = ARXIV:1302.5515;%%
  %4 citations counted in INSPIRE as of 09 Jan 2014

\bibitem{Lorce:2013bja} 
  C.~Lorc\'e,
  %``Gauge symmetry and background independence: Should the proton spin decomposition be path independent?,''
  Nucl.\ Phys.\ A {\bf 925}, 1 (2014).
  %[arXiv:1306.0456 [hep-ph]].
  %%CITATION = ARXIV:1306.0456;%%
  %6 citations counted in INSPIRE as of 24 Jun 2014

\bibitem{Lorce:2014mxa} 
  C.~Lorc\'e,
  %``Spin–orbit correlations in the nucleon,''
  Phys.\ Lett.\ B {\bf 735}, 344 (2014).
  %[arXiv:1401.7784 [hep-ph]].
  %%CITATION = ARXIV:1401.7784;%%
  %4 citations counted in INSPIRE as of 17 Sep 2014

\bibitem{Lorce:2015lna} 
  C.~Lorc\'e,
  %``The light-front gauge-invariant energy-momentum tensor,''
  arXiv:1502.06656 [hep-ph].
  %%CITATION = ARXIV:1502.06656;%%

\bibitem{Lorce:2011kd} 
  C.~Lorc\'e and B.~Pasquini,
  %``Quark Wigner Distributions and Orbital Angular Momentum,''
  Phys.\ Rev.\ D {\bf 84}, 014015 (2011).
  %[arXiv:1106.0139 [hep-ph]].
  %%CITATION = ARXIV:1106.0139;%%
  %42 citations counted in INSPIRE as of 09 Jan 2014

\bibitem{Lorce:2011zta} 
  C.~Lorc\'e and B.~Pasquini,
  %``On the Origin of Model Relations among Transverse-Momentum Dependent Parton Distributions,''
  Phys.\ Rev.\ D {\bf 84}, 034039 (2011).
  %[arXiv:1104.5651 [hep-ph]].
  %%CITATION = ARXIV:1104.5651;%%
  %19 citations counted in INSPIRE as of 17 Sep 2014

\bibitem{Lorce:2011kn} 
  C.~Lorc\'e and B.~Pasquini,
  %``Pretzelosity TMD and Quark Orbital Angular Momentum,''
  Phys.\ Lett.\ B {\bf 710}, 486 (2012).
  %[arXiv:1111.6069 [hep-ph]].
  %%CITATION = ARXIV:1111.6069;%%
  %11 citations counted in INSPIRE as of 26 Jun 2014

\bibitem{Lorce:2013pza} 
  C.~Lorc\'e and B.~Pasquini,
  %``Structure analysis of the generalized correlator of quark and gluon for a spin-1/2 target,''
  JHEP {\bf 1309}, 138 (2013).
  %[arXiv:1307.4497].
  %%CITATION = ARXIV:1307.4497;%%
  %8 citations counted in INSPIRE as of 26 Jun 2014

\bibitem{Lorce:2011dv} 
  C.~Lorc\'e, B.~Pasquini and M.~Vanderhaeghen,
  %``Unified framework for generalized and transverse-momentum dependent parton distributions within a 3Q light-cone picture of the nucleon,''
  JHEP {\bf 1105}, 041 (2011).
  %[arXiv:1102.4704 [hep-ph]].
  %%CITATION = ARXIV:1102.4704;%%
  %39 citations counted in INSPIRE as of 17 Sep 2014

\bibitem{Lorce:2011ni} 
  C.~Lorc\'e, B.~Pasquini, X.~Xiong and F.~Yuan,
  %``The quark orbital angular momentum from Wigner distributions and light-cone wave functions,''
  Phys.\ Rev.\ D {\bf 85}, 114006 (2012).
  %[arXiv:1111.4827 [hep-ph]].
  %%CITATION = ARXIV:1111.4827;%%
  %23 citations counted in INSPIRE as of 09 Jan 2014

\bibitem{Meissner:2009ww} 
  S.~Meissner, A.~Metz and M.~Schlegel,
  %``Generalized parton correlation functions for a spin-1/2 hadron,''
  JHEP {\bf 0908}, 056 (2009).
  %[arXiv:0906.5323 [hep-ph]].
  %%CITATION = ARXIV:0906.5323;%%
  %54 citations counted in INSPIRE as of 26 Jun 2014

\bibitem{Pasquini:2005dk}
  B.~Pasquini, M.~Pincetti and S.~Boffi,
  %``Chiral-odd generalized parton distributions in constituent quark models,''
  Phys.\ Rev.\  D {\bf 72}, 094029 (2005).
  %[arXiv:hep-ph/0510376];
  %%CITATION = PHRVA,D72,094029;%%

\bibitem{Pasquini:2006iv}
  B.~Pasquini, M.~Pincetti and S.~Boffi,
  %``Drell-Yan processes, transversity and light-cone wavefunctions,''
  Phys.\ Rev.\  D {\bf 76}, 034020 (2007).
  %[arXiv:hep-ph/0612094].
  %%CITATION = PHRVA,D76,034020;%%

\bibitem{Pasquini:2008ax}
  B.~Pasquini, S.~Cazzaniga, and S.~Boffi,
  %``Transverse momentum dependent parton distributions in a light-cone quark
  %model,''
  Phys.\ Rev.\  D {\bf 78}, 034025 (2008).
  %[arXiv:0806.2298 [hep-ph]].
  %%CITATION = PHRVA,D78,034025;%%

\bibitem{Penttinen:2000dg} 
  M.~Penttinen, M.~V.~Polyakov, A.~G.~Shuvaev and M.~Strikman,
  %``DVCS amplitude in the parton model,''
  Phys.\ Lett.\ B {\bf 491}, 96 (2000).
  %[hep-ph/0006321].
  %%CITATION = HEP-PH/0006321;%%
  %65 citations counted in INSPIRE as of 26 Jun 2014

\bibitem{Soper:1976jc} 
  D.~E.~Soper,
  %``The Parton Model and the Bethe-Salpeter Wave Function,''
  Phys.\ Rev.\ D {\bf 15}, 1141 (1977).
  %%CITATION = PHRVA,D15,1141;%%
  %191 citations counted in INSPIRE as of 22 Sep 2014

\bibitem{Stoilov:2010pv} 
  M.~N.~Stoilov,
  %``Hamiltonian analysis of the proposal by Chen et. al., PRL 100 (2008) 232002,''
  arXiv:1011.5617 [hep-th].
  %%CITATION = ARXIV:1011.5617;%%
  %1 citations counted in INSPIRE as of 03 Jun 2013

\bibitem{Wakamatsu:2010qj} 
  M.~Wakamatsu,
  %``On Gauge-Invariant Decomposition of Nucleon Spin,''
  Phys.\ Rev.\ D {\bf 81}, 114010 (2010).
  %[arXiv:1004.0268 [hep-ph]].
  %%CITATION = ARXIV:1004.0268;%%
  %54 citations counted in INSPIRE as of 09 Jan 2014

\bibitem{Wakamatsu:2010cb} 
  M.~Wakamatsu,
  %``Gauge and frame-independent decomposition of nucleon spin,''
  Phys.\ Rev.\ D {\bf 83}, 014012 (2011).
  %[arXiv:1007.5355 [hep-ph]].
  %%CITATION = ARXIV:1007.5355;%%
  %40 citations counted in INSPIRE as of 09 Jan 2014

\bibitem{Wakamatsu:2014zza} 
  M.~Wakamatsu,
  %``Is gauge-invariant complete decomposition of the nucleon spin possible?,''
  Int.\ J.\ Mod.\ Phys.\ A {\bf 29}, 1430012 (2014).
  %[arXiv:1402.4193 [hep-ph]].
  %%CITATION = ARXIV:1402.4193;%%
  %6 citations counted in INSPIRE as of 17 Sep 2014

\bibitem{Wakamatsu:2014toa} 
  M.~Wakamatsu,
  %``The problem of the gauge-invariant complete decomposition of the nucleon spin demystified,''
  arXiv:1409.4474 [hep-ph].
  %%CITATION = ARXIV:1409.4474;%%



\end{thebibliography}
\end{document}